\documentclass{article}
\usepackage{spconf,amsmath,graphicx,cite}
\usepackage{amsfonts}
\usepackage{booktabs}
\usepackage{url}
\usepackage{multicol}

\title{Improving speaker discrimination of target speech extraction with time-domain SpeakerBeam}
\name{Marc Delcroix$^1$, Tsubasa Ochiai$^1$, Katerina Zmolikova$^2$,  Keisuke Kinoshita$^1$,}
\secondlinename{ Naohiro Tawara$^1$, Tomohiro Nakatani$^1$, Shoko Araki$^1$}

\address{$^1$NTT Communication Science Laboratories, NTT Corporation, Kyoto, Japan \\
  $^2$Brno University of Technology, Speech@FIT and IT4I Center of Excellence, Czechia}%
\begin{document}
\ninept
\maketitle
\begin{abstract}
Target speech extraction, which extracts a single target source in a mixture given clues about the target speaker, has attracted increasing attention.
We have recently proposed SpeakerBeam, which exploits an adaptation utterance of the target speaker to extract his/her voice characteristics that are then used to guide a neural network towards extracting speech of that speaker.
SpeakerBeam presents a practical alternative to speech separation as it enables tracking speech of a target speaker across utterances, and achieves promising speech extraction performance. However, it sometimes fails when speakers have similar voice characteristics,  such as in same-gender mixtures, because it is difficult to discriminate the target speaker from the interfering speakers.
In this paper, we investigate strategies for improving the speaker discrimination capability of SpeakerBeam. 
First, we propose a time-domain implementation of SpeakerBeam similar to that proposed for a time-domain audio separation network (TasNet), which has achieved state-of-the-art performance for speech separation. 
Besides, we investigate (1) the use of spatial features to better discriminate speakers when microphone array recordings are available, (2) adding an auxiliary speaker identification loss for helping to learn more discriminative voice characteristics.
We show experimentally that these strategies greatly improve speech extraction performance, especially for same-gender mixtures, and outperform TasNet in terms of target speech extraction.
\end{abstract}
\begin{keywords}
Target speech extraction, time-domain network, spatial features, multi-task loss
\end{keywords}
\section{Introduction}
\label{sec:introduction}
Recently, deep learning based speech separation approaches have attracted increasing attention\cite{hershey2016deep,kolbaek2017multitalker,kinoshita2018listening,luo2018tasnet}.
Earlier approaches such as deep clustering~\cite{hershey2016deep} and permutation invariant training (PIT)~\cite{kolbaek2017multitalker}, performed processing in the frequency-domain and generated time-frequency masks for each source in the mixture. More recently, a convolutional time-domain audio separation network (Conv-TasNet) has been proposed and led to great separation performance improvement surpassing ideal time-frequency masking~\cite{luo2018tasnet,luo2019conv,shi2019furcax}. 
The separation performance of TasNet has been further improved by exploiting spatial information when a microphone array is available~\cite{bahmaninezhad2019comprehensive}.

Despite the great success of neural network-based speech separation, it requires knowing or estimating the number of sources in the mixture and still suffers from a global permutation ambiguity issue, i.e. an arbitrary mapping between source speakers and outputs. These limitations arguably limit the practical usage of speech separation. 
In contrast to speech separation, target speech extraction exploits an auxiliary clue to identify the target speaker in the mixture and extracts only speech of that speaker. After our initial work~\cite{zmolikova2017spkaware,zmolikova2019Journal}, research on target speech extraction has then gained increasing attention~\cite{Chen2018DeepEN,LookingToListen2018,Wang_voicefilter19,ChenglinXuIcassp19,GuanjunIS19}, as it naturally avoids the global permutation ambiguity issue and does not require knowing the number of sources in the mixtures.

We have proposed SpeakerBeam~\cite{zmolikova2017spkaware,zmolikova2019Journal}, which is a target speech extraction method that exploits a speaker embedding vector derived from an adaptation or enrollment utterance of the target speaker to guide a neural network towards extracting speech of that speaker.
This is realized by combining two networks, a sequence summary network~\cite{vesely2016sequence} that computes the speaker embedding vector from the amplitude spectrum of the adaptation utterance and a speech extraction network that accepts the amplitude spectrum of the speech mixture and the embedding vector as inputs and generates a time-frequency mask for extracting the target speaker. In this paper, we call this approach frequency-domain SpeakerBeam (FD-SpeakerBeam).

We have shown that FD-SpeakerBeam could achieve competitive speech extraction performance and be used as a front-end for automatic speech recognition (ASR)~\cite{zmolikova2019Journal}. However, we observe a great performance gap between same-gender and different-gender mixtures~\cite{delcroixIcassp19}. 
It is indeed difficult to discriminate a target speaker in a mixture when speakers have similar voice characteristics.

In this paper, we investigate strategies to tackle this issue.
First, following the success of TasNet, we propose \emph{a time-domain implementation of SpeakerBeam} (TD-SpeakerBeam), whose speech extraction network accepts time-domain signals of the mixture, and outputs directly the time-domain signal of the target speaker.
We also replace the sequence summary network with a convolutional network to obtain richer speaker embedding vectors.

Moreover, to further improve speaker discrimination capability, we extend TD-SpeakerBeam to accept spatial information from microphone array recordings as additional input features. 
We argue that simply adding spatial features to the input of TD-SpeakerBeam may limit the potential to process spatial information.
Consequently, we propose an alternative approach, called \emph{internal combination}, for exploiting spatial information more effectively within the SpeakerBeam framework. 

Finally,  to enforce learning more discriminative speaker embedding vectors, we propose using a multi-task loss for training SpeakerBeam, that combines a speech reconstruction loss with a \emph{speaker identification loss} (SI-loss).

We performed experiments on two datasets, which show that 
(1) TD-SpeakerBeam greatly improves target speech extraction performance and outperforms a competitive system based on TasNet separation followed by an x-vector~\cite{SnyderSLT16} based target speech selection module, (2) exploiting spatial features with the proposed internal combination helps target speaker extraction especially for same-gender mixtures, (3) the additional SI-loss consistently improves performance when a sufficient number of speakers are included in the training data, (4) by varying the number of training speakers, although TasNet performance does not change significantly, SpeakerBeam benefits greatly from more training speakers especially for same-gender mixtures because it helps improving target speakers identification.
These results confirm the efficiency of the proposed strategies for improving the target speaker discrimination capability of SpeakerBeam.

\section{Proposed time-domain SpeakerBeam}

Let us first describe the implementation of TD-SpeakerBeam.
Then, in section \ref{ssec:IPD}, we  discuss  approaches for exploiting spatial 
information when microphone array recordings are available.
Finally, in section \ref{ssec:SIL}, we introduce a multi-task loss to improve speaker discrimination even when only a single microphone is available.
\begin{figure}[t]
  \centering
  \centerline{\includegraphics[width=8cm]{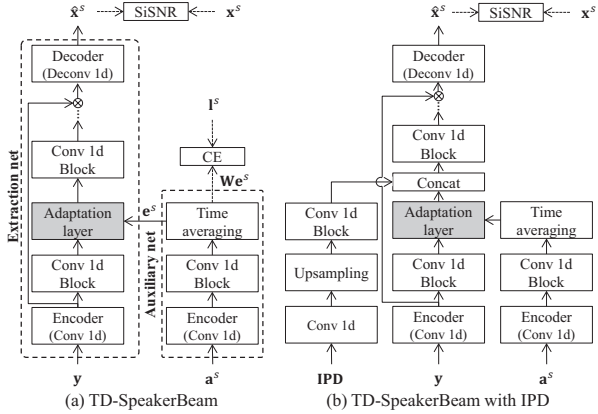}}
\vspace{-2mm}
\caption{Block diagram of (a) proposed TD-SpeakerBeam and (b) TD-SpeakerBeam with internal combination of IPD features.}
\label{fig:diagram}
\vspace{-0.5cm}
\end{figure}

\subsection{TD-SpeakerBeam}
Figure \ref{fig:diagram}-(a) is a block diagram of the proposed TD-SpeakerBeam.
Let $\mathbf{y}$, $\mathbf{a}^s$ and $\hat{\mathbf{x}}^s$ be the time-domain signals of the speech mixture, the adaptation utterance, and the estimated target speech for target speaker $s$.
SpeakerBeam is composed of two networks, an \emph{extraction network}, and an \emph{auxiliary network}.
In the original FD-SpeakerBeam~\cite{zmolikova2019Journal}, these networks accept the amplitude spectrum of the mixture and adaptation signals as inputs and generate a time-frequency mask. In this paper, we modify the implementation of these networks to input and output time-domain signals. 

The time-domain extraction network follows a similar configuration as Conv-TasNet~\cite{luo2019conv}, i.e. it consists of a 1d convolution layer that accepts the mixture signal $\mathbf{y}$ (encoder layer), 
several convolution blocks, 
and finally, a 1d deconvolution (decoder layer) that outputs the extracted speech signal in the time-domain, $\hat{\mathbf{x}}^s$.

There are two major differences with Conv-TasNet. First, the output consists of a single signal corresponding to the target speech only.
Second, we insert an adaptation layer between the first and second convolution blocks\footnote{We found in preliminary experiments that placing the adaptation layer after the first convolution block achieved the best performance.}
to drive the network towards extracting the target speaker. 
The adaptation layer accepts a speaker embedding vector of the target speaker, $\mathbf{e}^s$, as auxiliary information. We use a multiplicative adaptation layer following our previous work~\cite{delcroixIcassp19}, although other adaptation layers could be used~\cite{zmolikova2019Journal,ChenglinXuIcassp19}.

The target speaker embedding vector, $\mathbf{e}^s$, is computed by the time-domain auxiliary network. In the original FD-SpeakerBeam, the auxiliary network consists of a sequence summary network~\cite{vesely2016sequence}, i.e. a few fully connected layers followed by a time-averaging operation. 
Here, we propose using a convolutional auxiliary network to accept the time-domain input signal of the adaptation utterance $\mathbf{a}^s$. The auxiliary network consists of an encoder layer and a single convolution block similar to those used in the extraction network. 

\subsection{Spatial features}
\label{ssec:IPD}
Spatial information extracted from multi-channel recordings can provide an alternative source of information about the mixtures that could help discriminate speakers better.
There have been several works showing the benefit of adding spatial features to the input of speech enhancement networks~\cite{ArakiMCDAE2015,WangMCDeepClustering18,bahmaninezhad2019comprehensive}.
For example, \cite{bahmaninezhad2019comprehensive} recently showed that the inter-microphone phase difference (IPD) features could improve the separation performance of TasNet in reverberant conditions.
The IPD of the mixture signal between two microphones is defined as,
\begin{equation}
    \Phi_{i,j,t,f} = \angle\left( \frac{Y_{i,t,f}}{Y_{j,t,f}}\right),
\end{equation}
where $Y_{i,t,f} \in \mathbb{C}$ is the short-time Fourier transform (STFT) coefficient of the mixture signal at microphone $i$,
$t$ is the time-frame index, and $f$ is the frequency index.
Here we limit our investigation to the two-microphone case. 
Following \cite{bahmaninezhad2019comprehensive}, we use cosine and sine of the IPDs as spatial features,
\begin{eqnarray}
\mathbf{IPD}_t&=&[\cos(\Phi_{1,2,t,1}), \ldots, \cos(\Phi_{1,2,t,F}), \nonumber \\
&&\sin(\Phi_{1,2,t,1}), \ldots, \sin(\Phi_{1,2,t,F})],
\end{eqnarray}
where $F$ is the number of frequency bins.
Note that the frame size and window shift of the STFT used to compute the IPD features may differ from the window size and shift used in the encoder of the extraction network. IPD features are thus upsampled to match the settings of the extraction network

IPD features provide spatial information related to the direction of sources in the mixture.
SpeakerBeam extracts the target speaker based on the speaker embedding vector, $\mathbf{e}^s$, that may represent ``spectral'' information\footnote{Strictly speaking it is not the usual spectrum information as we use a learnable convolutional encoder layer to analyze the signal instead of STFT.} about the target speaker, but does not include spatial information. Consequently, it is not obvious how to efficiently combine the IPD features and the target speaker embedding vector as they represent different information.
In this paper, we consider two schemes, \emph{input combination} and \emph{internal combination}.

The \emph{input combination} is similar to that proposed for TasNet in \cite{bahmaninezhad2019comprehensive}, where the IPD features (processed with a convolutional layer and upsampled) are concatenated to the output of the encoder layer of the extraction network. 
Input combination may force the initial convolution block to combine spatial information from the IPD features and ``spectral'' information from the mixture signal $\mathbf{y}$ into a ``spectral'' representation, which allows the adaptation layer (coming after the first convolutional block) to select the target speaker by comparing this ``spectral'' representation with the target speaker embedding vector, $\mathbf{e}^s$. This may reduce the potential of the network to fully exploit spatial information by the upper layers of the network.

Figure \ref{fig:diagram}-(b) shows a schematic diagram of TD-SpeakerBeam with the alternative \emph{internal IPD combination}. 
It combines the IPD features (processed with a 1D convolutional layer, upsampling, and a convolution block) after the adaptation layer. 
Therefore, this lets the speaker selection operate based only on the ``spectral'' information,
and the spatial information can be exploited by the upper layers without being obstructed by the adaptation layer.

Here, we only consider exploiting spatial information as additional information to the extraction network.
Besides, SpeakerBeam can also be combined with beamforming, which is particularly efficient for ASR applications~\cite{zmolikova2017spkaware,ochiai_ICASSP2020}, but is out of the scope of this paper.
 
\subsection{multi-task learning with additional SI-loss}
\label{ssec:SIL}
The extraction network and auxiliary networks are trained jointly from random initialization given the speech mixtures, $\mathbf{y}$, adaptation utterances, $\mathbf{a}^s$, and the target speech signals $\mathbf{x}^s$. In our previous works~\cite{zmolikova2017spkaware,zmolikova2019Journal}, we trained SpeakerBeam using only a target speech reconstruction loss.
In this paper, we propose using a multi-task loss for training TD-SpeakerBeam that combines scale-invariant source-to-noise ratio (SiSNR)~\cite{roux2019sdr} as signal reconstruction loss and cross-entropy-based SI-loss.
The SI-loss is used to obtain more discriminative speaker embedding vectors.
The multi-task loss is given by,
\begin{equation}
    L(\Theta|\mathbf{y},\mathbf{a}^s,\mathbf{x}^s, \mathbf{l}^s)= -\text{SiSNR}(\mathbf{x}^s,\hat{\mathbf{x}}^s) + \alpha \text{CE}(\mathbf{l}^s, \sigma(\mathbf{W}\mathbf{e}^s)) ,
    \label{eq:loss}
\end{equation}
where $\Theta$ are the model parameters, and $\mathbf{l}^s$ is a one-hot vector representing the target speaker ID, $\text{SiSNR}(\mathbf{x}^s,\hat{\mathbf{x}}^s)$ is the SiSNR between the estimated and true target speech, 
$\text{CE}(\mathbf{l}^s, \sigma(\mathbf{W}\mathbf{e}^s))$ is the cross entropy between the speaker label $\mathbf{l}^s$ 
and the speaker embedding vector projected onto the training speaker space, $\mathbf{W}\mathbf{e}^s$,  $\mathbf{W}$ is a projection matrix,  $\sigma(\cdot)$ is a softmax function, and $\alpha$ is a scaling parameter.

\section{Related prior work}
An alternative way to perform target speech extraction consists of performing speech separation followed by target speaker selection from the separated signals. Such a scheme was proposed in~\cite{DrudeIcassp18} for deep attractor network~\cite{chen2017deep}, but to the best of our knowledge has not been investigated for time-domain separation approaches. In the experiments, we compare TD-SpeakerBeam with TasNet separation followed by x-vector-based speaker selection~\cite{SnyderSLT16}, which can be considered a strong baseline for target speech extraction.

We borrowed from previous works on multi-channel source separation~\cite{WangMCDeepClustering18,bahmaninezhad2019comprehensive} that IPD features may be good candidates for increasing extraction performance. 
Besides adding IPD features to the extraction network, an alternative approach was recently proposed~\cite{GuanjunIS19}, where a set of fixed beamformers combined with an attention module on the output of the beamformers was used to perform a rough initial target speech extraction followed by a refinement step with FD-SpeakerBeam. 
Investigating such a scheme with TD-SpeakerBeam or other spatial features will be part of our future works. 

\section{Experiments}
\begin{table}[t]
  \renewcommand{\arraystretch}{1.0}
  \caption{Amount of data and number of female (F) and male (M) speakers.}
%\vspace{2mm}
  \label{tab:corpora}
  \centering
  %\scalebox{0.9}{
  \begin{tabular}{l c@{ }c@{ }c@{  }c c@{ }c@{ }c@{ }c}
    \toprule
    	&	\multicolumn{4}{c}{Train} &	\multicolumn{4}{c}{Test} \\
    & \#Mixtures & \#Spks & \#F & \#M & \#Mixtures & \#Spks & \#F & \#M  \\
    \midrule
    WSJ &20k & 101 & 52 & 49 & 3k & 18 & 8 & 10 \\    
    \midrule
    CSJ & 50k & 937 & 166 & 771  & 15k & 30 & 10 & 20 \\	
    \bottomrule
  \end{tabular}
  
\vspace{-3mm}
\end{table}

\subsection{Datasets}
We performed experiments using two datasets, multi-channel WSJ0 2 mixtures (MC-WSJ0-2 mix) and CSJ-2mix.
Table \ref{tab:corpora} shows details of the amount of data and the number of female and male speakers in the training and test sets. 

MC-WSJ0-2 mix is a publicly available multi-channel version of the WSJ0-2mix corpus \cite{wang2018multi} that consists of mixtures of WSJ0 utterances\cite{wsj}. Multi-channel recordings are generated by convolving clean speech signals with room impulse responses simulated with the image method for reverberation time of up to about 600ms. The dataset consists of 8 channel recordings, but we use only 2 channels in our experiments. This dataset has only 101 training speakers. We use it thus only for the investigations on the use of spatial features.

The second dataset consists of single-channel 2-speaker mixtures that we simulated by mixing utterances from the corpus of spontaneous Japanese (CSJ)\cite{Maekawa_CSJ2000} at SNR between -5 and 5 dBs. This dataset has a larger number of training speakers (937 speakers) and is used to investigate the effect of the SI-loss and the impact of the number of training speakers. 

For both datasets, we randomly selected adaptation utterances of the target speaker in a mixture from the utterances of that speaker that differed from the utterance in the mixture.
In the MC-WSJ0-2mix experiments, the adaptation utterances did not contain reverberation, although a similar level of performance could be achieved with reverberant adaptation utterances. We used an 8kHz sampling frequency for all our experiments.

\subsection{Experimental settings}
TD-SpeakerBeam was implemented based on the open source Conv-TasNet implementation~\cite{funcwj}. 
In particular, following the hyper-parameter notations in the original paper~\cite{luo2019conv}, 
we set the hyper-parameters to N=256, L=20, B=256, H=512, P=3, X=8, R=4.
The auxiliary network consisted of an encoder layer and a single convolution block.

We compare the proposed TD-SpeakerBeam with 
(1) TasNet with oracle target speech selection,
(2) TasNet with x-vector-based target speech selection, and  
(3) our previous implementation of FD-SpeakerBeam.
TasNet used the network configuration described in~\cite{luo2019conv,funcwj}, with hyper-parameters equivalent to those of TD-SpeakerBeam. 
Oracle speaker selection was performed by finding the speaker permutation that maximizes the signal-to-distortion ratio (SDR).
For x-vector-based speaker selection~\cite{SnyderSLT16}, we selected the target speech as the output of the TasNet separation module whose x-vector presented the highest cosine similarity with the x-vector of the adaptation utterance. We used the publicly available x-vector extractor model that was trained on multi-condition noisy and reverberant data to extract x-vectors~\cite{sellDieHard18,dihard19}.

The network architecture of FD-SpeakerBeam consisted of 3 BLSTM layers followed by a sigmoid layer and 3 fully connected layers for the auxiliary network.
FD-SpeakerBeam was trained with the MSE loss between the amplitude spectrum of clean target speech and masked signals.
Details of the configuration can be found in~\cite{zmolikova2019Journal}. 
Note that many aspects of the network configuration and the training procedure differ from that of TD-SpeakerBeam.
Consequently, the results of FD-SpeakerBeam are only indicative of the level of performance achieved in our previous works. A more fair comparison between the impact of working in the time and frequency domain in the context of speech separation can be found in~\cite{bahmaninezhad2019comprehensive}.

For experiments with MC-WSJ0-2mix, we extracted IPD features using an STFT window of 32 msec and a shift of 16 msec. 
We compare TasNet with input IPD combination~\cite{bahmaninezhad2019comprehensive} and TD-SpeakerBeam with input and internal IPD combination.

All time-domain models were trained to optimize the SiSNR criterion only (i.e. $\alpha=0$ in Eq.~(\ref{eq:loss})) 
except when we mentioned the use of the SI-loss, in which case we used $\alpha=10$. 
As the evaluation metrics, we used the scale-invariant SDR of BSSeval~\cite{vincent2006performance}
 
\subsection{Results with IPD features using MC-WSJ0-2mix}

\begin{table}[t]
  \renewcommand{\arraystretch}{1.0}
  \caption{SDR (dB) on the MC-WSJ0-2mix corpus. Bold-fonts indicate best performance (except for oracle).}
%\vspace{2mm}
  \label{tab:result_wsj0}
  \centering
  \scalebox{0.9}{
  \begin{tabular}{l@{ }l c c c c c}
    \toprule
    &	&IPD	&	FF&	MM&	FM&	avg \\
    \midrule
    (1) &Mixture             &	-	 & 0.17	 & 0.16	& 0.16	& 0.16 \\
    \midrule
    (2) &TasNet (oracle)  &	-   &		\emph{8.68}	& \emph{9.75}	& \emph{12.14}	& \emph{10.84}\\
	(3) &                    & input	&  \emph{11.52}	& \emph{11.37}	& \emph{12.17}	& \emph{11.83}\\
	(4) &TasNet (xvect)  &   -   &  4.59	& 4.93	& 11.44	& 8.35\\
	(5) &                    & input	&  6.01	& 5.80	& 11.35	& 8.80\\
    \midrule
    (6)  &FD-SpkBeam &	-	&  5.19	& 5.32	& 10.27	& 7.94\\
    \midrule
    (7) & TD-SpkBeam	&   -	&  9.13	& 9.47	& \bf{12.77}	& 11.17\\
	(8) &                    & input	 &	9.02 & 	9.71	& 12.55	& 11.11\\
	(9) &                    & internal   &\bf{10.17}	& \bf{10.30}	& 12.49	&  \bf{11.45}\\
    \bottomrule
  \end{tabular}
  }
\vspace{-5mm}
\end{table}

Table \ref{tab:result_wsj0} shows the SDR for the MC-WSJ0-2mix experiments for mixtures of female-female (FF), male-male (MM) and female-male (FM) speakers.
We confirmed that TasNet with oracle target speaker selection (row (2)) achieved high SDR performance.
Moreover, TasNet with input combination of IPD features (row (3)) further improved performance especially for mixtures of same-gender speakers. 
These results can be considered an upper-bound for TasNet-based target speaker extraction.
We omitted results with the internal IPD combination for TasNet, as it performed worse than using IPD features at the input. 

Performance dropped greatly when using x-vector-based speaker selection (row (4) and (5)), especially for FF and MM cases. 
Although the x-vector extractor was trained on multi-condition data, it may still be affected by reverberation, which may partly contribute to the poor performance.
However, reverberation is not the only reason for the performance drop because x-vector selection performed significantly worse than oracle even for the following CSJ experiments that do not include reverberation.

FD-SpeakerBeam (row (6)) performed slightly worse than TasNet (xvect).
The proposed TD-SpeakerBeam (row (7)) outperformed all systems but TasNet with oracle speaker selection. 
Especially, there is a smaller performance gap between mixtures of speakers of the same and different genders than with FD-SpeakerBeam. 
We further confirmed that TD-SpeakerBeam with internal IPD combination (row (9)) improved performance by up to 1 dB and performed better than input combination (row (8)).

\subsection{Results with the SI-loss on CSJ-2mix}

\begin{table}[t]
  \renewcommand{\arraystretch}{1.0}
  \caption{SDR [dB] on the CSJ-2mix corpus.}
%\vspace{2mm}
  \label{tab:results_csj}
  \centering
  \scalebox{0.9}{
  \begin{tabular}{l@{ }l c c c c }
    \toprule
    &                        & FF	& MM	& FM	& avg \\
    \midrule
(1) &Mixture			            & 0.19	& 0.18	& 0.18	& 0.18 \\
    \midrule
(2) &TasNet (oracle)			& \emph{11.86}	& \emph{14.81}	& \emph{17.01}	& \emph{15.28} \\
(3) &TasNet (xvect)			& 7.65	& 12.51	& 16.29	& 13.35 \\
    \midrule
(4) &FD-SpkBeam (Freq)			    & 6.42	& 8.35	& 10.52	& 8.93 \\
    \midrule
(5) &TD-SpkBeam			    & 12.56	& 17.15	& 18.83	& 17.24 \\
(6) &TD-SpkBeam + SI-loss     	& \bf{13.60}	& \bf{17.75}	& \bf{19.22}	& \bf{17.81}\\
    \bottomrule
  \end{tabular}
  }
\vspace{-5mm}
\end{table}

\begin{figure}[tb]
\begin{minipage}[t]{0.5\linewidth}
  %\centering
  \centerline{\includegraphics[width=4.3cm]{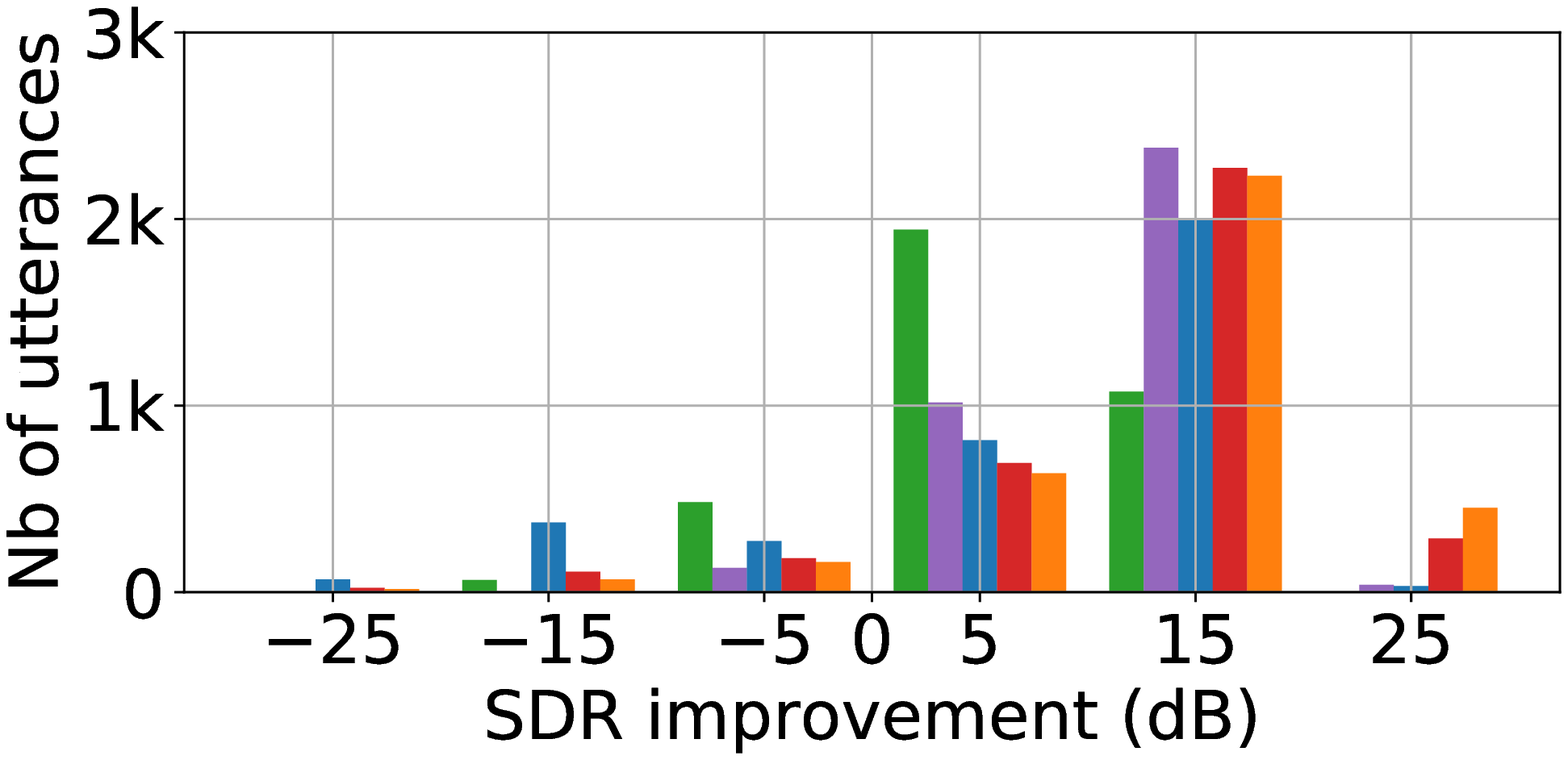}}
  \centerline{(a) FF}%\medskip
%  \vspace{2.0cm}
\end{minipage}
%\vfill%
\begin{minipage}[t]{0.5\linewidth}
    \centerline{\includegraphics[width=4.3cm]{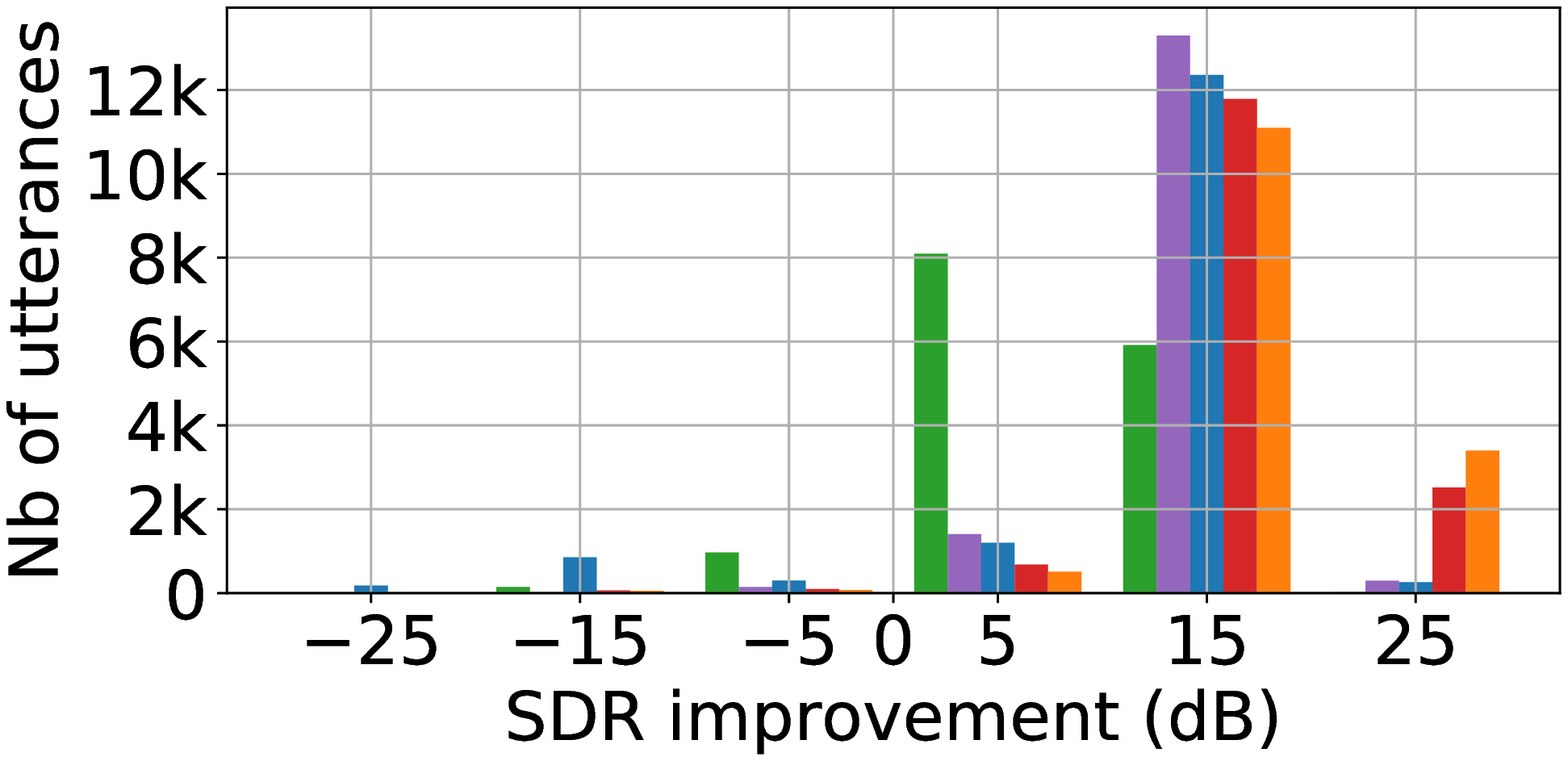}}
%  \vspace{1.5cm}
  \centerline{(b) MM}%\medskip
\end{minipage} \\

\begin{minipage}[b]{1\linewidth}
  \centering
  \centerline{\includegraphics[width=6.5cm]{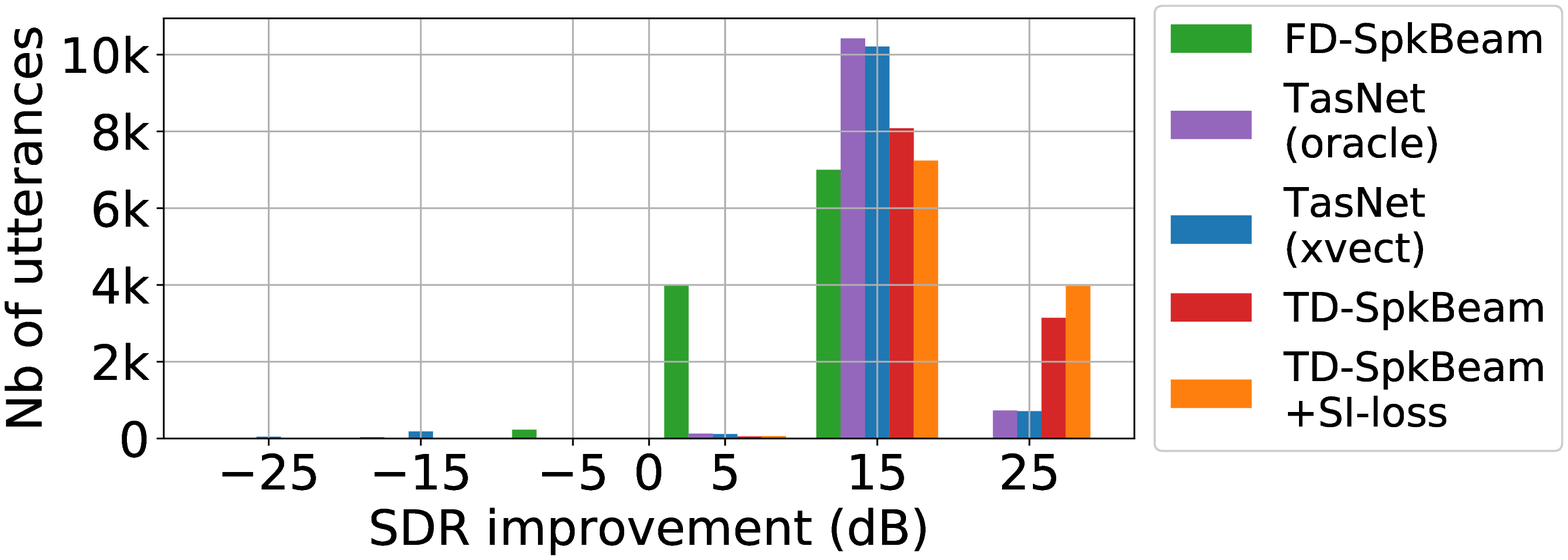}}
%  \vspace{1.5cm}
  \centerline{(c) FM}%\medskip
	%\vfill
%\begin{minipage}[b]{0.5\linewidth}
  \centering
\end{minipage}
\vspace{-7mm}
\caption{Histogram of the SDR improvement for CSJ 2 mix experiments with 937 training speakers.}
\label{fig:res_csj_distirbution}
%\vspace{-0.5cm}
\end{figure}

Table \ref{tab:results_csj} shows the SDR for  TasNet with oracle and x-vector-based speaker selection, FD-SpeakerBeam and TD-SpeakerBeam without and with SI-loss.
These results were obtained when using all 937 training speakers for training the models. 
TD-SpeakerBeam achieved much better performance than FD-SpeakerBeam 
and TasNet with or without oracle speaker selection. 
Moreover, SI-loss provided further consistent performance improvement of up to 1 dB.

Figure \ref{fig:res_csj_distirbution} shows the histogram of the SDR improvement for FF, MM and FM mixtures. 
TD-SpeakerBeam with or without SI-loss greatly reduced processing failures (SDR improvement of 0 dB or less).
Moreover, the SI-loss led to better overall performance (more results with high SDR improvement).

\begin{figure}[tb]
\begin{minipage}[t]{0.3\linewidth}
  %\centering
  \centerline{\includegraphics[width=2.7cm]{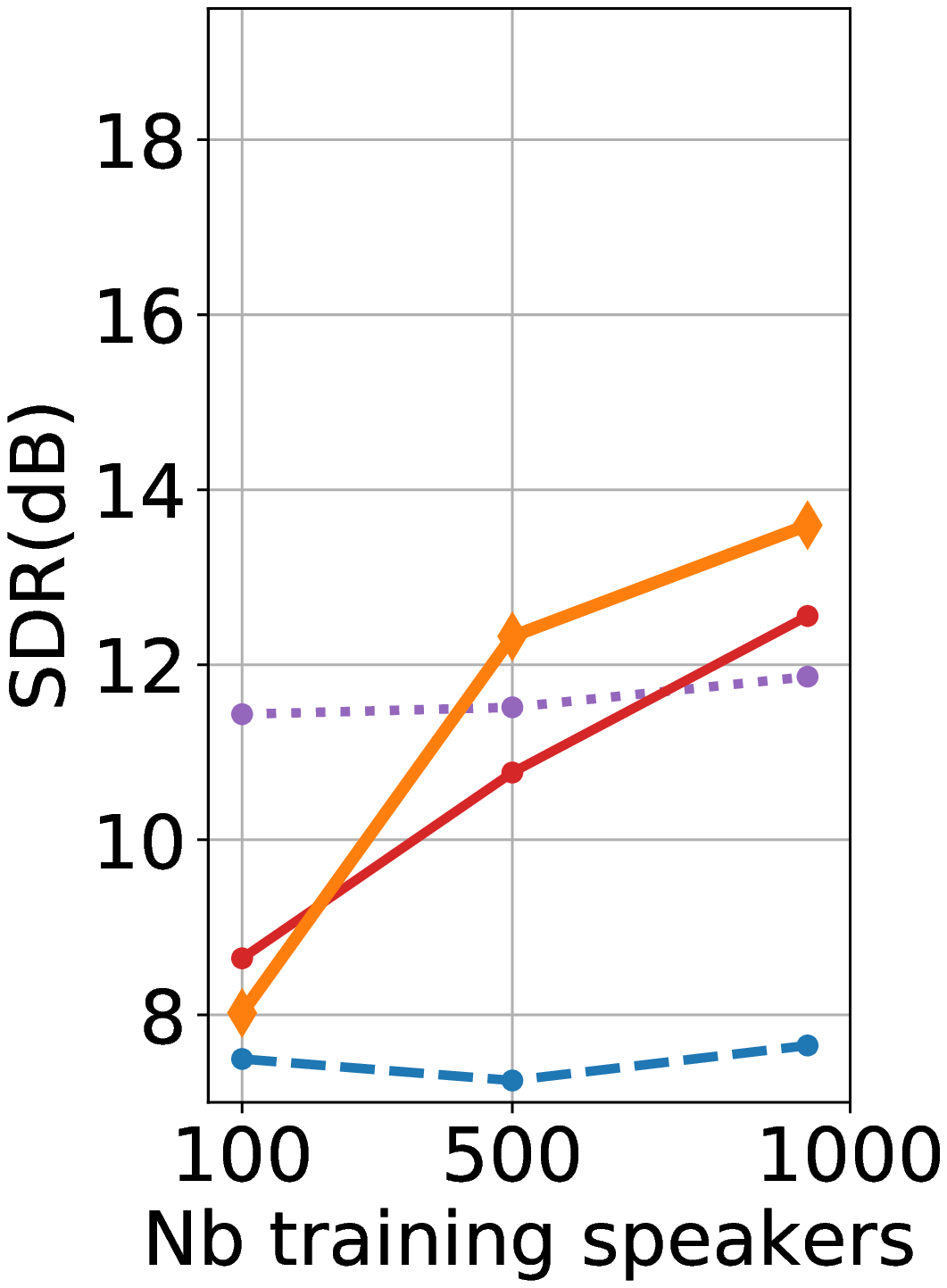}}
%  \vspace{2.0cm}
  \centerline{(a) FF}%\medskip
\end{minipage}
%\vfill%
\begin{minipage}[t]{0.3\linewidth}
  
    \centerline{\includegraphics[width=2.7cm]{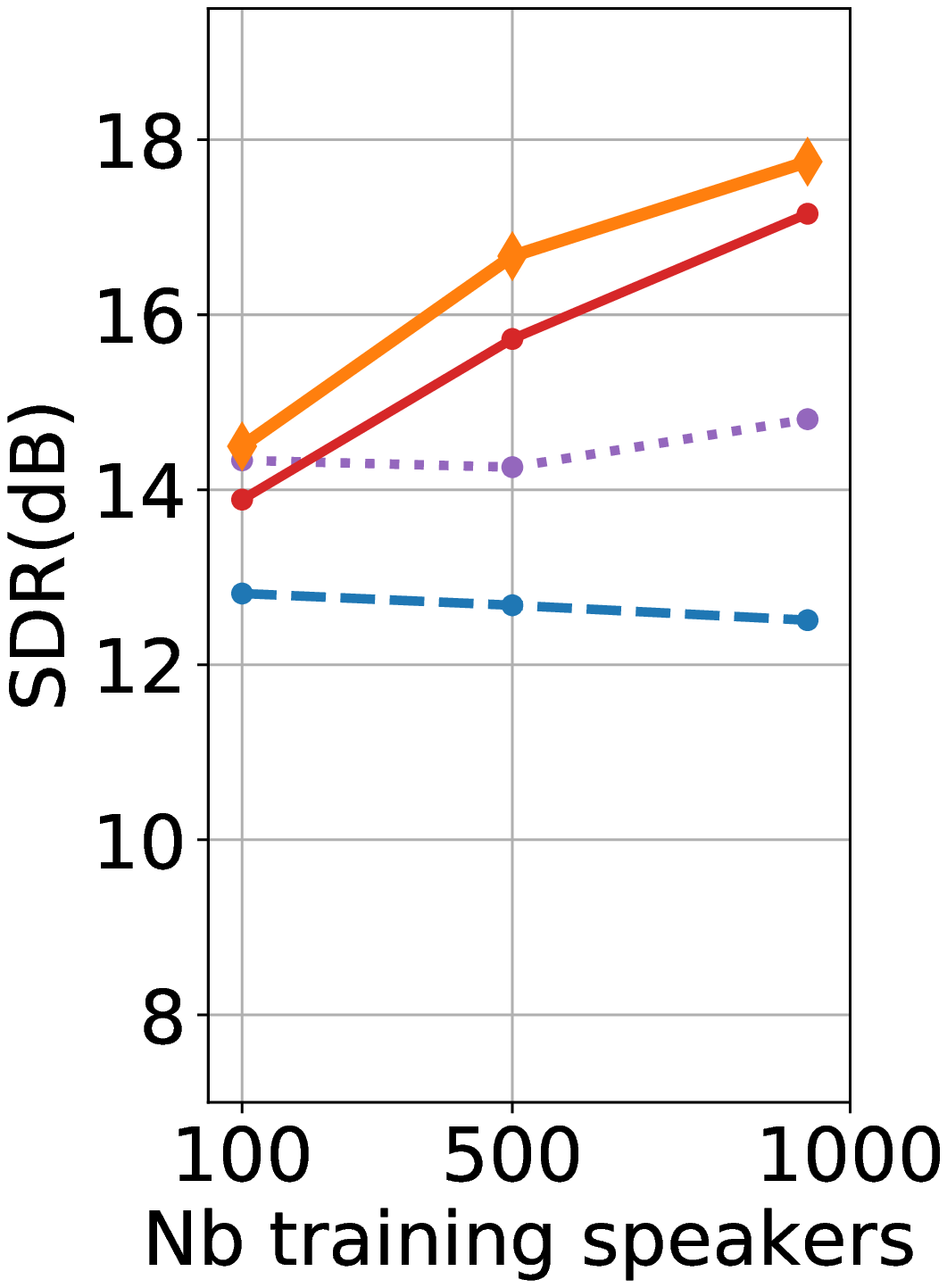}}
%  \vspace{1.5cm}
  \centerline{(b) MM}%\medskip
\end{minipage}
%\vfill%
\begin{minipage}[t]{0.3\linewidth}
  %\centering
%\end{minipage}
%\vfill
%\begin{minipage}[b]{0.5\linewidth}
  \centering
  \centerline{\includegraphics[width=2.7cm]{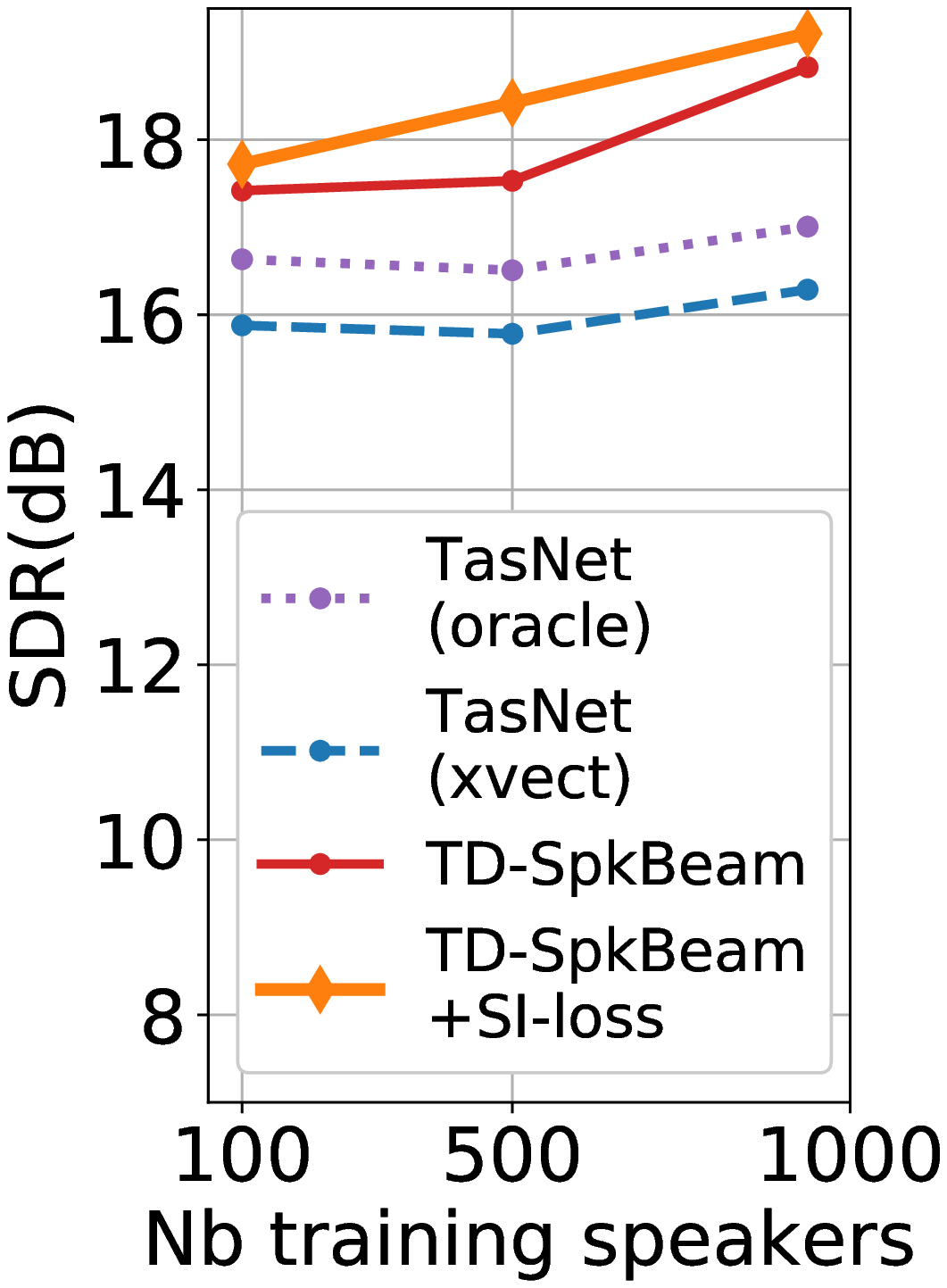}}
%  \vspace{1.5cm}
  \centerline{(c) FM}%\medskip
	%\vfill
%\begin{minipage}[b]{0.5\linewidth}
  \centering
\end{minipage}
\vspace{-2mm}
\caption{SDR as a function of the number of training speakers.}
\label{fig:res_csj_spk_nb}
\vspace{-0.5cm}
\end{figure}

Figure \ref{fig:res_csj_spk_nb} shows the SDR as a function of the number of training speakers.
The curves were obtained by creating 3 different training sets with 100, 500 and all 937 training speakers.
In all cases, we used 50k mixtures. % so that the amount of training data stays roughly the same. 
Interestingly, we observe that increasing the number of speakers has little effect on TasNet performance, 
but greatly improved the performance of SpeakerBeam. This suggests that, for SpeakerBeam, separating signals is somewhat easier than identifying speakers. The SI-loss provided consistent improvement when using more than 100 speakers (This is why we did not use the SI-loss in the MC-WSJ0 experiments).
Note that performance remains significantly lower for FF mixtures, partly because there are fewer female speakers in the training set (see table \ref{tab:corpora}) and also because it appears to be more challenging to separate FF mixtures, as also suggested by the lower performance of TasNet in this case.
\section{Conclusion}
\label{sec:conclusion}
In this paper, we proposed different strategies for improving the target speech discrimination capability of SpeakerBeam.
We showed that a time-domain implementation greatly improved performance.
Moreover, the performance gap between same-gender and different-gender mixtures could be reduced further by exploiting spatial information, using an additional SI-loss, or by increasing the number of training speakers.

In future work, we would like to combine these techniques to tackle more challenging noisy and reverberant mixtures, e.g. \cite{delcroixIcassp19}.
Moreover, we will also investigate other approaches to integrate spatial information \cite{bahmaninezhad2019comprehensive,GuanjunIS19} and more discriminative SI-losses~\cite{XiangIS19}.
\vfill\pagebreak
\clearpage

\bibliographystyle{IEEEbib}
\bibliography{strings,refs}

\begin{thebibliography}{10}

\bibitem{hershey2016deep}
J.~R. Hershey, Z. Chen, J. Le~Roux, and S. Watanabe,
\newblock ``Deep clustering: Discriminative embeddings for segmentation and
  separation,''
\newblock in {\em Proc. of ICASSP'16}, 2016, pp. 31--35.

\bibitem{kolbaek2017multitalker}
M. Kolb{\ae}k, D. Yu, Z.-H. Tan, and J. Jensen,
\newblock ``Multitalker speech separation with utterance-level permutation
  invariant training of deep recurrent neural networks,''
\newblock {\em IEEE/ACM Trans. ASLP}, vol. 25, no. 10, pp. 1901--1913, 2017.

\bibitem{kinoshita2018listening}
K. Kinoshita, L. Drude, M. Delcroix, and T. Nakatani,
\newblock ``Listening to each speaker one by one with recurrent selective
  hearing networks,''
\newblock in {\em Proc. of ICASSP'18}, 2018, pp. 5064--5068.

\bibitem{luo2018tasnet}
Y. Luo and N. Mesgarani,
\newblock ``{TasNet}: Surpassing ideal time-frequency masking for speech
  separation,''
\newblock in {\em Proc. of ICASSP'18}, 2018.

\bibitem{luo2019conv}
Y. Luo and N. Mesgarani,
\newblock ``{Conv-TasNet}: Surpassing ideal time--frequency magnitude masking
  for speech separation,''
\newblock {\em IEEE/ACM Trans. ASLP}, vol. 27, no. 8, pp. 1256--1266, 2019.

\bibitem{shi2019furcax}
Z. Shi, H. Lin, L. Liu, R. Liu, S. Hayakawa, and J. Han,
\newblock ``Furcax: End-to-end monaural speech separation based on deep gated
  (de) convolutional neural networks with adversarial example training,''
\newblock in {\em Proc. of ICASSP'19}, 2019, pp. 6985--6989.

\bibitem{bahmaninezhad2019comprehensive}
F. Bahmaninezhad, J. Wu, R. Gu, S.-X. Zhang, Y. Xu, M. Yu, and D. Yu,
\newblock ``A comprehensive study of speech separation: spectrogram vs waveform
  separation,''
\newblock in {\em Proc. of Interspeech'19}, 2019, pp. 4574--4578.

\bibitem{zmolikova2017spkaware}
K. Zmolikova, M. Delcroix, K. Kinoshita, T. Higuchi, A. Ogawa, and T. Nakatani,
\newblock ``Speaker-aware neural network based beamformer for speaker
  extraction in speech mixtures,''
\newblock in {\em Proc. of Interspeech'17}, 2017, pp. 2655--2659.

\bibitem{zmolikova2019Journal}
K. Zmolikova, M. {Delcroix}, K. {Kinoshita}, T. {Ochiai}, T. {Nakatani}, L.
  {Burget}, and J. {Cernocky},
\newblock ``{SpeakerBeam}: Speaker aware neural network for target speaker
  extraction in speech mixtures,''
\newblock {\em IEEE Journal of Selected Topics in Signal Processing}, vol. 13,
  no. 4, pp. 800--814, 2019.

\bibitem{Chen2018DeepEN}
J. Chen, D. Su, L. Chen, M. Yu, Y. Qian, and D. Yu,
\newblock ``Deep extractor network for target speaker recovery from single
  channel speech mixtures,''
\newblock in {\em Proc. of Interspeech'18}, 2018, pp. 307--311.

\bibitem{LookingToListen2018}
A. Ephrat, I. Mosseri, O. Lang, T. Dekel, K. Wilson, A. Hassidim, W.~T.
  Freeman, and M. Rubinstein,
\newblock ``Looking to listen at the cocktail party: A speaker-independent
  audio-visual model for speech separation,''
\newblock {\em ACM Trans. on Graphics}, vol. 37, no. 4, pp. 112:1--112:11,
  2018.

\bibitem{Wang_voicefilter19}
Q. Wang, H. Muckenhirn, K. Wilson, P. Sridhar, Z. Wu, J.~R. Hershey, R.~A.
  Saurous, R.~J. Weiss, Y. Jia, and I.~L. Moreno,
\newblock ``{VoiceFilter}: Targeted voice separation by speaker-conditioned
  spectrogram masking,''
\newblock in {\em Proc. of Interspeech'19}, 2019, pp. 2728--2732.

\bibitem{ChenglinXuIcassp19}
C. Xu, W. Rao, E.~S. Chng, and H. Li,
\newblock ``Optimization of speaker extraction neural network with magnitude
  and temporal spectrum approximation loss,''
\newblock in {\em Proc. of ICASSP’19}, 2019, pp. 6990--6994.

\bibitem{GuanjunIS19}
G. Li, S. Liang, S. Nie, W. Liu, M. Yu, L. Chen, S. Peng, and C. Li,
\newblock ``Direction-aware speaker beam for multi-channel speaker
  extraction,''
\newblock in {\em Proc. of Interpseech'19}, 2019.

\bibitem{vesely2016sequence}
K. Vesely, S. Watanabe, K. Zmolikova, M. Karafiat, L. Burget, and J.~H.
  Cernocky,
\newblock ``Sequence summarizing neural network for speaker adaptation,''
\newblock in {\em Proc. of ICASSP'16}, 2016, pp. 5315--5319.

\bibitem{delcroixIcassp19}
M. Delcroix, K. Zmolikova, T. Ochiai, K. Kinoshita, S. Araki, and T. Nakatani,
\newblock ``Compact network for {SpeakerBeam} target speaker extraction,''
\newblock in {\em Proc. of ICASSP’19}, 2019, pp. 6965--6969.

\bibitem{SnyderSLT16}
D. {Snyder}, P. {Ghahremani}, D. {Povey}, D. {Garcia-Romero}, Y. {Carmiel}, and
  S. {Khudanpur},
\newblock ``Deep neural network-based speaker embeddings for end-to-end speaker
  verification,''
\newblock in {\em Proc. of SLT'16}, 2016, pp. 165--170.

\bibitem{ArakiMCDAE2015}
S. {Araki}, T. {Hayashi}, M. {Delcroix}, M. {Fujimoto}, K. {Takeda}, and T.
  {Nakatani},
\newblock ``Exploring multi-channel features for denoising-autoencoder-based
  speech enhancement,''
\newblock in {\em Proc. of ICASSP'15}, 2015, pp. 116--120.

\bibitem{WangMCDeepClustering18}
Z. {Wang}, J. {Le Roux}, and J.~R. {Hershey},
\newblock ``Multi-channel deep clustering: Discriminative spectral and spatial
  embeddings for speaker-independent speech separation,''
\newblock in {\em Proc. of ICASSP'18}, 2018.

\bibitem{ochiai_ICASSP2020}
T. Ochiai, M. Delcroix, R. Ikeshita, K. Kinoshita, T. Nakatani, and S. Araki,
\newblock ``Beam-{TasNet}: Time-domain audio separation network meets
  frequency-domain beamformer,''
\newblock in {\em Proc. of ICASSP'20 (Submitted)}, 2020.

\bibitem{roux2019sdr}
J. Le~Roux, S. Wisdom, H. Erdogan, and J.~R. Hershey,
\newblock ``{SDR}-half-baked or well done?,''
\newblock in {\em Proc. of ICASSP'19}, 2019, pp. 626--630.

\bibitem{DrudeIcassp18}
L. {Drude}, T. {von Neumann}, and R. {Haeb-Umbach},
\newblock ``Deep attractor networks for speaker re-identification and blind
  source separation,''
\newblock in {\em Proc. of ICASSP'18}, 2018, pp. 11--15.

\bibitem{chen2017deep}
Z. Chen, Y. Luo, and N. Mesgarani,
\newblock ``Deep attractor network for single-microphone speaker separation,''
\newblock in {\em Proc. of ICASSP'18}, 2017, pp. 246--250.

\bibitem{wang2018multi}
Z.-Q. Wang, J. Le~Roux, and J.~R. Hershey,
\newblock ``Multi-channel deep clustering: Discriminative spectral and spatial
  embeddings for speaker-independent speech separation,''
\newblock in {\em Proc. of ICASSP'18}, 2018, pp. 1--5.

\bibitem{wsj}
J. Garofolo,
\newblock ``{CSR-I (WSJ0) Complete LDC93S6A},''
  \url{https://catalog.ldc.upenn.edu/ldc93s6a}, 1993.

\bibitem{Maekawa_CSJ2000}
K. Maekawa, H. Koiso, S. Furui, and I. H.,
\newblock ``Spontaneous speech corpus of {J}apanese,''
\newblock in {\em Proc. of LREC'00}, 2000, pp. 947--952.

\bibitem{funcwj}
``{\footnotesize \url{https://github.com/funcwj/conv-tasnet}},'' .

\bibitem{sellDieHard18}
G. Sell, D. Snyder, A. McCree, D. Garcia-Romero, J. Villalba, M. Maciejewski,
  V. Manohar, N. Dehak, D. Povey, S. Watanabe, and S. Khudanpur,
\newblock ``Diarization is hard: Some experiences and lessons learned for the
  {JHU} team in the inaugural {DIHARD} challenge,''
\newblock in {\em Proc. of Interspeech'18}, 2018, pp. 2808--2812.

\bibitem{dihard19}
``{\footnotesize \url{https://github.com/iiscleap/DIHARD-2019-baseline}},'' .

\bibitem{vincent2006performance}
E. Vincent, R. Gribonval, and C. F{\'e}votte,
\newblock ``Performance measurement in blind audio source separation,''
\newblock {\em IEEE trans. ASLP}, vol. 14, no. 4, pp. 1462--1469, 2006.

\bibitem{XiangIS19}
X. Xiang, S. Wang, H. Huang, Y. Qian, and K. Yu,
\newblock ``Margin matters: Towards more discriminative deep neural network
  embeddings for speaker recognition,''
\newblock in {\em Proc. of Interpseech'19}, 2019.

\end{thebibliography}

\end{document}